\documentclass[twocolumn,prd,preprintnumbers,amsmath,amssymb,nofootinbib]{revtex4}

\usepackage{graphicx}
\usepackage{dcolumn}
\usepackage{bm}
\usepackage{epsfig}
\usepackage{epstopdf}
\usepackage{amssymb}
\usepackage{amsmath}
\usepackage[utf8]{inputenc}

\newcommand{\beq}{\begin{equation}}
\newcommand{\eeq}{\end{equation}}

\newcommand{\rhor}{\rho_{r}}
\newcommand{\rhop}{\rho_{\phi}}
\newcommand{\rhoc}{\rho_{\chi}}

\newcommand{\Gp}{\Gamma_\phi}

\newcommand{\ad}{a_D}
\newcommand{\vd}{v_D}

\newcommand{\sigv}{\langle \sigma v \rangle}
\newcommand{\B}{\gamma_D v_D}
\newcommand{\arh}{a_{\mathrm{RH}}}
\newcommand{\trh}{T_{\mathrm{RH}}}
\newcommand{\aeq}{a_{\mathrm{eq}}}
\newcommand{\khm}{k_{\mathrm{hm}}}
\newcommand{\Mhm}{M_{\mathrm{hm}}}

\begin{document}

\title{Constraining Nonthermal Dark Matter's Impact on the Matter Power Spectrum}

\author{Carisa Miller}
\email{carisa@live.unc.edu}
\author{Adrienne L. Erickcek}
\email{erickcek@physics.unc.edu}
\affiliation{Department of Physics and Astronomy, University of North Carolina at Chapel Hill, Phillips Hall CB3255, Chapel Hill, NC 27599 USA}
\author{Riccardo Murgia}
\email{riccardo.murgia@sissa.it}
\affiliation{SISSA, Via Bonomea 265, 34136 Trieste, Italy}
\affiliation{INFN, Sezione di Trieste, via Valerio 2, 34127 Trieste, Italy}
\affiliation{IFPU, Institute for Fundamental Physics of the Universe, via Beirut 2, 34151 Trieste, Italy}


\begin{abstract}

The inclusion of a period of (effective) matter domination following inflation and prior to the onset of radiation domination has interesting and observable consequences for structure growth. During this early matter-dominated era (EMDE), the Universe was dominated by massive particles, or an oscillating scalar field, that decayed into Standard Model particles, thus reheating the Universe.  This decay process could also be the primary source of dark matter. In the absence of fine-tuning between the masses of the parent and daughter particles, both dark matter particles and Standard Model particles would be produced with relativistic velocities. We investigate the effects of the nonthermal production of dark matter particles with relativistic velocities on the matter power spectrum by determining the resulting velocity distribution function for the dark matter. We find that the vast majority of dark matter particles produced during the EMDE are still relativistic at reheating, so their free streaming erases the perturbations that grow during the EMDE. The free streaming of the dark matter particles can also prevent the formation of satellite galaxies around the Milky Way and the structures observed in the Lyman-$\alpha$ forest. For a given reheat temperature, these observations put an upper limit on the velocity of the dark matter particles at their creation. For example, for a reheat temperature of 10 MeV, dark matter must be produced with a Lorentz factor $\gamma \lesssim 550$.

\end{abstract}

\maketitle
\section{Introduction}
\label{sec:Intro}

The nature of dark matter remains a pressing question in cosmology. One of the most common assumptions is that dark matter was once in thermal equilibrium with Standard Model (SM) particles in the early Universe. As the SM plasma cooled, thermal production of dark matter ceased while annihilations continued. The dark matter abundance thus began decreasing until its annihilation rate equaled the Hubble rate, at which point annihilations also ceased, and the dark matter abundance became constant. A second common assumption is that this dark matter freeze-out process occurred during a period of radiation domination. These assumptions allow one to calculate the annihilation rate that generates the currently observed dark matter abundance. The required annihilation cross section is ``miraculously'' of the electroweak scale \cite{historyofDM}. However, as we continually place more stringent bounds on dark matter properties, while failing to receive signals from any direct \cite{LUX,XENON,Panda} or indirect \cite{indirect1,indirect2,indirect3,indirect4,indirect5,indirect6} searches, interest in alternatives to this commonly considered scenario grows.

Alternatives to the common scenario often challenge the assumptions that dark matter was in thermal equilibrium with SM particles and that it froze out during an era of radiation domination, both of which, while tenable, are not strictly necessary. A period of radiation domination is required at temperatures below $\sim\!\!3$ MeV in order to be consistent with the successful predictions of light element abundances from Big Bang Nucleosynthesis (BBN)\cite{3mev1,3mev2,3mev3}. Inflation, however, is believed to occur at energy scales that greatly exceed this temperature, and the thermal history of the Universe between the two periods is entirely unconstrained. In the simplest scenario, the inflaton decays into relativistic particles that come to dominate the energy density of the Universe, and an era of radiation domination begins \cite{infl1,infl2}.  The transition to a radiation-dominated era, known as reheating, is usually assumed to occur at temperatures many orders of magnitude above \mbox{$3$ MeV}. It is not necessary, however, that this be the case - the reheating of the Universe can occur at any temperature between $3$ MeV and the energy scale of inflation, and it can be caused by a number of different mechanisms.

In many models, inflation ends when the scalar field that drives inflation begins oscillating in its potential minimum before decaying. If these oscillations occur in a quadratic potential, the field behaves as pressureless fluid, and the Universe is effectively matter dominated \cite{pressureless}. Similar scenarios occur when one considers the scalar (moduli) fields that are a common component of string theories \cite{moduli1,moduli2,moduli3,moduli4,moduli5,moduli6,moduli7,moduli8}. These oscillating fields naturally come to dominate the energy density of the Universe following the decay of the inflaton, providing another viable mechanism to produce an effectively matter-dominated era. Hidden-sector theories, in which the dark matter does not couple directly to the SM, can also alter the thermal history \cite{hidden,hidden2,hidden3,hidden4,TommiHS2,TommiHS1}, providing yet another means to achieve a period of matter domination prior to BBN. Thus, an early matter-dominated era (EMDE) arises in many theories of the early Universe.

The occurrence of an EMDE can profoundly affect dark matter phenomenology, notably its resulting relic abundance \cite{ra1,ra4,ra2,ra3,GG,GGshort,ra6,ra5,ra7,Faz2,Pankaj1,Pankaj2,Tommi2,Tommi1,Faz1,Chowdhury,DiMarco}. The entropy generated by the decay of the dominant matter component during the EMDE dilutes the relic abundance of existing particles; if dark matter thermally decoupled during the EMDE, a smaller annihilation cross section $\sigv$ is required to compensate for this dilution and provide the observed dark matter abundance.  Contrarily, if dark matter is a decay product of the dominating component, its abundance can be significantly enhanced, requiring a larger $\sigv$ to compensate for the excess, a scenario already under pressure by $\gamma$-ray observations \cite{winoveritas,wino}. The correct relic abundance can almost always be obtained with the appropriate combinations of $\sigv$, dark matter branching ratio, and temperature at reheating \cite{GG,GGshort,Faz2,Faz1}.  In many scenarios, the dominant production mechanism for dark matter is by decay, rather than thermal production.

Another interesting consequence of an EMDE is the growth of small-scale structure. Subhorizon density perturbations in dark matter grow linearly with the scale factor during an EMDE, as opposed to the much slower logarithmic growth experienced during a radiation-dominated era \cite{Adrienne,fan,Adrienne2015}. This linear growth can provide an enhancement to dark matter structure on extremely small scales ($\lambda \lesssim 30$ pc for temperature at reheating $> 3$ MeV), providing observable consequences to this scenario if dark matter is a cold thermal relic \cite{Adrienne2015,Adrienne2016,newhooper}.

However, if the dark matter is relativistic at reheating, the perturbation modes that enter the horizon during the EMDE will be wiped out by the free streaming of dark matter particles \cite{Adrienne,fan}. For this reason, Ref. \cite{Adrienne} assumed that the dark matter particles were born from the decay process with nonrelativistic velocities or had a way of rapidly cooling in order for the enhancement to substructure to be preserved. Assuming a nonrelativistic initial velocity for the dark matter requires a small, finely tuned mass splitting between the parent and daughter particles, and it is more natural to assume any daughter particles are produced relativistically.

Reference \cite{fan} claimed that the large free-streaming length of dark matter produced relativistically from scalar decay would washout any enhancement to structure growth. However, Ref. \cite{fan} reached this conclusion by assuming that all dark matter particles were created at reheating, neglecting those particles created during the EMDE. The momenta of particles born prior to reheating decreased throughout the EMDE. Consequently, particles born earlier will be slower at reheating. We investigate the extent to which the redshifting of the particles' momenta affects their velocity distribution at reheating, focusing on the average particle velocity and the fraction of particles below a given velocity, to determine under what conditions the EMDE enhancement to structure growth can be preserved.

We further consider under what conditions the free streaming of relativistically produced dark matter could suppress the structures we observe. The Lyman-$\alpha$ forest provides information on the matter power spectrum at the smallest observable scales, $0.5 \mathrm{Mpc}/h < \lambda < 20 \mathrm{Mpc}/h$ \cite{lyman1,lyman2,lyman3}. The Milky Way's (MW) satellite galaxies also constrain the small-scale power spectrum \cite{Milkway}. Preventing the suppression of power at these scales provides us with constraints on the allowed dark matter velocity at its production for a given reheat temperature.

This paper is organized as follows. We begin in Section \ref{sec:1} by introducing our model for reheating and nonthermal dark matter production and the resulting evolution of the average dark matter velocity. In Section \ref{sec:dmdf} we derive a distribution function for the dark matter and use it to examine the fraction of dark matter that is nonrelativistic at reheating and the fraction whose velocity is sufficiently low to preserve the EMDE-enhanced structure formation. In Section \ref{sec:lyalpha} we examine conditions under which the dark matter velocity is high enough to run afoul of constraints from Lyman-$\alpha$ forest observations and observed MW satellites.  We conclude in Section \ref{sec:end}. Throughout this paper we will use natural units: $c=\hbar=k_B=1$.

\section{Nonthermal Production of Dark Matter}
\label{sec:1}

In the scenario we consider, the energy density of the Universe is dominated by an oscillating scalar field (or a massive particle species). As previously mentioned, for sufficiently rapid oscillations within a quadratic potential, the field's energy density scales as $\rhop \propto a^{-3}$, and it exhibits the same dynamics and perturbation evolution as a pressureless fluid \cite{pressureless,lingrow2,lingrow1}. The Universe experiences an early ``matter''-dominated era until the expansion rate equals the decay rate of the field, $H\simeq\Gp$, at which point the Universe transitions from scalar to radiation domination. We use this transition to define the reheat temperature, $T_{\mathrm{RH}}$:
\begin{align}\label{eq:deftrh}
\sqrt{\frac{4\pi^3 G}{45}g_{*,\mathrm{RH}}T_{\mathrm{RH}}^4} &= \Gamma_\phi,
\end{align}
where $G$ is the gravitational constant and $g_{*,\mathrm{RH}}$ is the number of relativistic degrees of freedom at $T_{\mathrm{RH}}$. For a scalar field that decays into both dark matter and relativistic particles, the equations for the evolution of the energy densities of the scalar field $\rhop$, relativistic SM particles $\rhor$, and dark matter $\rhoc$ are given by
\begin{align}
\label{eq:energydens}
\dot{\rho}_{\phi} =& - 3H\rhop -\Gp\rhop,\\ \nonumber
\dot{\rho}_r =& - 4H\rhor + (1-f)\Gp\rhop,\\ \nonumber
\dot{\rho}_{\chi} =& - 3H(1+w_\chi)\rhoc + f\Gp\rhop.
\end{align}
Here dots represent differentiation with respect to proper time, $f$ is the fraction of the scalar's energy that is transferred to the dark matter, and $w_\chi$ is the dark matter equation-of-state parameter.

\subsection{Dark Matter Abundance}
\label{sec:abundance}

In the above system of equations, we do not allow for scattering interactions between the dark matter and SM particles. Also, we neglect both the thermal production and self-annihilation of dark matter particles, effectively assuming that the velocity-averaged annihilation cross section is small enough that any amount of dark matter lost to annihilations is negligible and any produced thermally is negligible compared to that produced from scalar decay. However, if dark matter annihilations are $s$-wave, neglecting annihilations does not change the results of our conclusions because we are interested in the \textit{average} dark matter velocity and the \textit{fraction} of dark matter that has lost sufficient momentum to participate in structure formation. These quantities are dependent on the velocity distribution of dark matter. For $s$-wave annihilations, the velocity-averaged cross section is independent of particle velocity, and the distribution of particle velocities would be unaffected by the inclusion of annihilations.

Without annihilations, constraining the reheat temperature to be above 3 MeV, as required by BBN, leads to a direct constraint on the fraction of the scalar's energy imparted to the dark matter: $f \lesssim 10^{-7}$ \cite{Adrienne} for nonrelativistic dark matter. This branching ratio is quite small and it would be more natural to expect the energy imparted in the decay of the scalar to be more evenly allocated to both the dark matter and the SM.  The inclusion of annihilations, however, significantly reduces the ratio of dark matter to radiation. This can allow for a more balanced transfer of energy, $f\sim0.5$, while still achieving a sufficiently small dark matter abundance through annihilations \cite{fan,GG}.  For relativistic dark matter, its abundance is also dependent on the velocity imparted to the particles at decay $(v_D)$.

\begin{figure}
\centering\includegraphics[width=3.4in]{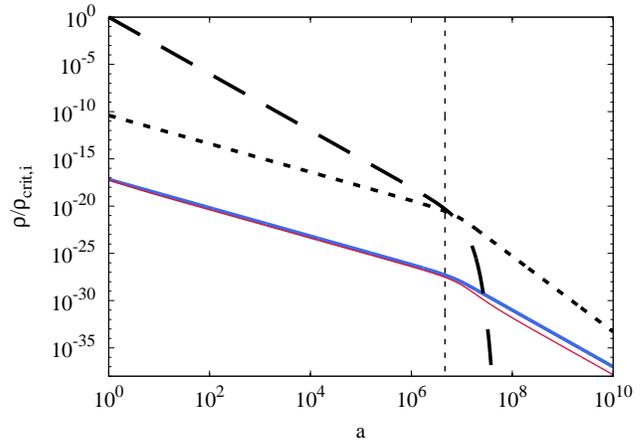}
\caption{The energy densities of the scalar (black, solid), radiation (grey, solid), and dark matter for particles born relativistic ($v_D = 0.99$; red, dotted) and nonrelativistic ($v_D = 0.1$; blue, dashed). During the EMDE, both the dark matter and radiation densities scale as $a^{-3/2}$ while they are being sourced by the decaying scalar field. Here $f=10^{-7}$ and the scalar decay rate is $\tilde{\Gamma}_\phi = \Gamma_\phi / H_i = 10^{-10}$. Reheating is marked by the thin vertical dashed line at $\arh = \tilde{\Gamma}_\phi^{-2/3}$.}
\label{rhos}
\end{figure}

In our model, we consider the scenario in which the dark matter is produced via a two-body decay so that all dark matter particles are born with the same velocity, $v_D$. The energy density and decay rate of the scalar then govern the evolution of the equation of state $w_\chi$ for the dark matter particles during the EMDE. The rates at which new particles are produced and the momentum of existing particles redshifts away determine the average energy per particle of the dark matter:
\begin{align}\label{eq:avgenergy}
\langle E \rangle = \frac{\int_{1}^{a}\sqrt{m_\chi^2 + (\gamma m_\chi v(a,\ad))^2}\frac{d\hat{n}_\chi}{d\ad}d\ad}{\int_{1}^{a}\frac{d\hat{n}_\chi}{d\ad}d\ad},
\end{align}
where $\hat{n}_\chi$ is the comoving number density of the dark matter particles, $v(a,\ad)$ is the velocity at $a$ of a dark matter particle created at $\ad$, and we integrate over $\ad$ with $a=1$ setting the onset of dark matter production.

When evaluating Eq. (\ref{eq:avgenergy}) we use the fact that the comoving number density of the dark matter evolves according to
\begin{align}\label{}
\frac{d\hat{n}_\chi}{dt} = b \Gamma_\phi \hat{n}_\phi,
\end{align}
where $\hat{n}_\phi$ is the comoving number density of $\phi$ particles, and $b$ is the number of dark matter particles produced per scalar decay. We can then (following a procedure similar to that in Ref. \cite{twobody}) express the term $d\hat{n}_\chi/da_D$ as
\begin{align}\label{eq:dnda}
\frac{d\hat{n}_\chi}{da_D} = \frac{d\hat{n}_\chi}{dt}\frac{dt}{da_D} = \frac{b \Gamma_\phi \hat{n}_\phi}{\dot{a}_D} =   \frac{b \Gamma_\phi \frac{\rho_\phi}{m_\phi}a_D^3}{a_D H_D} \propto \frac{\rho_\phi}{H_D}a_D^2.
\end{align}
The constants $b$, $\Gamma_\phi$, and $m_\phi$ appear in both integrals in Eq. (\ref{eq:avgenergy}) and consequently do not affect $\langle E \rangle$. We numerically evaluate Eq. (\ref{eq:avgenergy}) to obtain the average energy as a function of the scale factor; this is made even simpler by noting that the contribution of the dark matter to the expansion rate at the time of decay, $H_D$, is entirely negligible compared to both the scalar and radiation energy densities. The calculation of the average energy then informs how the dark matter equation of state evolves:
\begin{align}\label{eq:wchi}
w_\chi = - \frac{1}{3H\langle E \rangle} \frac{d\langle E \rangle}{dt}.
\end{align}
The mass of the dark matter particle can be pulled from both the average energy and its derivative, and so $w_\chi$ at any given time depends only on the average velocity.

Using Eq. (\ref{eq:wchi}), we numerically solve the set of equations in Eq. (\ref{eq:energydens}) with the initial condition $a_i \equiv a(t_i) = 1$, and we assume there is no dark matter in existence prior to this time. Figure \ref{rhos} shows the evolution of the scalar, radiation, and relativistic $(v_D = 0.99)$ and nonrelativistic $(v_D = 0.1)$ dark matter energy densities in our model. The energy densities in the figure are given as fractions of the initial critical energy density $\rho_{\mathrm{crit},i}$.

In Fig. \ref{rhos}, we have chosen to fix $f= 10^{-7}$, which directly sets the relative abundance of dark matter to radiation during the EMDE to be $\sim \!\! 10^{-7}$, and we have chosen a scalar decay rate $\tilde{\Gamma}_{\phi} \equiv \Gamma_\phi / H_i = 10^{-10}$, which sets $\arh \equiv \tilde{\Gamma}^{-2/3} \simeq 5\times 10^6$. It is worth noting that, by this definition, $\arh \neq a |_{T=\trh}$. The numerical solutions to Eq. (\ref{eq:energydens}) show that by a scale factor of $3\arh$ enough of the scalar has decayed away that it is a negligible source of radiation. As a result, the radiation energy density then evolves as in the usual radiation-dominated era from a temperature $T(3\arh) \simeq 0.34 T_{\mathrm{RH}}$ onward. This sets the relation between $\arh$ and $\trh$ to be $\arh / a_0 = 1.54 (T_0 / \trh) g_{*S}^{-1/3}(0.34\trh)$, where $g_{*S}$ is the number of relativistic degrees of freedom in the entropy density.

The effects of the dark matter particles' velocities can already been seen in Fig. \ref{rhos}. During the EMDE, the energy density of any species, relativistic or nonrelativistic, sourced by scalar decay evolves as $\rho\propto a^{-3/2}$. At the end of the EMDE, when the scalar field is no longer sourcing new particles, the energy densities of the decay products will begin to scale as $\rho\propto a^{-3(w+1)}$. For a scenario in which dark matter is born relativistic, the average value of $w_\chi$ during reheating is close to $1/3$, and the dark matter behaves more like radiation. In Fig.~\ref{rhos} we can see that, following reheating, the energy density of relativistic dark matter (dotted) redshifts away faster than its nonrelativistic counterpart (dashed). Once the scalar field has decayed completely and there is no creation of new, hot particles, the existing particles' momenta continue to redshift until the average particle is no longer relativistic, and after that, the dark matter density scales as $a^{-3}$.

Increasing the velocity imparted to the dark matter particles upon their creation increases the time it takes after reheating for the dark matter energy density to begin scaling as $a^{-3}$, thus increasing the duration of radiation domination for a given value of $f$. The temperature at matter-radiation equality is $T_{\mathrm{eq}} = 0.796 \pm 0.005$ eV \cite{Planck}, and so a longer radiation-dominated era implies a higher temperature at reheating. For a fixed value of $f$, the reheat temperature that matches the observed dark matter abundance in a scenario of relativistically produced dark matter is a factor of $\gamma_D$ greater than that of nonrelativistic dark matter, where $\gamma_D$ is the Lorentz factor of the relativistic dark matter particle at production. For a given reheat temperature, the value of $f$ required to produce the observed dark matter abundance is
\begin{align}\label{eq:foftrh}
f \simeq 2.3 \times 10^{-7} (3 \mathrm{MeV} / T_{\mathrm{RH}}) \gamma_D.
\end{align}
In the absence of annihilations, the dark matter density during the EMDE is also determined by $f$: if $v_D \ll 1$ then $\rho_\chi / \rho_r \simeq (5/3) f$ \cite{Adrienne}, whereas if $v_D \simeq 1$ then $\rho_\chi / \rho_r \simeq f$. In the latter case, $\rho_\chi / \rho_r \simeq f$ continues until the dark matter is no longer relativistic or changes in $g_*$ disrupt the $a^{-4}$ scaling of $\rho_r$. Therefore, for $\gamma_D \gg 1$, obtaining the observed dark matter abundance requires that $\rho_\chi / \rho_r \simeq 2.3 \times 10^{-7} (3 \mathrm{MeV} / T_{\mathrm{RH}}) \gamma_D$ shortly after reheating.  This requirement applies regardless of whether or not annihilations alter the dark matter abundance during the EMDE. After reheating, but while the dark matter is still relativistic, the relative dark matter abundance will evolve as
\begin{align}\label{eq:pr/px}
\left.\frac{\rho_\chi}{\rho_r}\right|_T \simeq \ & 2.3 \times 10^{-7} \left(\frac{3 \mathrm{MeV}}{ T_{\mathrm{RH}}}\right) \gamma_D \\ \nonumber
& \times \left[\frac{g_{*S}(T)}{g_{*S}(0.34\trh)}\right]^{4/3} \frac{g_*(0.34\trh)}{g_*(T)},
\end{align}
in either case.

If dark matter is still relativistic at neutrino decoupling, it could affect the predictions of BBN. Dark matter produced relativistically at reheating would still be relativistic ($\gamma \gtrsim 2$) when \mbox{$T= 10$ MeV} if
\begin{align}\label{}
\gamma_D \gtrsim 2.4 g_{*S}^{1/3}(0.34\trh) \frac{\trh}{10 \mathrm{MeV}}.
\end{align}
Relativistic dark matter behaves as an additional radiation component, and can be characterized as a change in the number of effective neutrinos, $\Delta N_\mathrm{eff}$. The energy density in relativistic particles can be written as
\begin{align}\label{}
\rho_r + \rho_{\chi,\mathrm{rel}} & = \frac{\pi^2}{30}\left[g_* + \frac{7}{8} \times 2 \times \Delta N_\mathrm{eff} \left(\frac{T_\nu}{T}\right)^4 \right] T^4, \\ \nonumber
\rho_{\chi,\mathrm{rel}} & = \frac{\pi^2}{30}\left(\frac{7}{8} \times 2 \times \Delta N_\mathrm{eff} \right) T_\nu^4,
\end{align}
where $T_\nu$ is the neutrino temperature, which we assume evolves as $a^{-1}$ after $T=T_\nu=10$ MeV, when $g_* = g_{*S} =10.75$. Thus, the fractional component of the energy density in dark matter at 10 MeV is related to $\Delta N_\mathrm{eff}$ by
\begin{align}\label{}
f_{10\mathrm{MeV}} \equiv \left.\frac{\rho_\chi}{\rho_r}\right|_{10\mathrm{MeV}} & = \frac{7}{4} \Delta N_\mathrm{eff} \ g_{*S}^{-1}(10 \mathrm{MeV}) \nonumber \\
&\simeq 0.074 \Delta N_\mathrm{eff}.
\end{align}
Bounds from BBN constrain $N_\mathrm{eff} = 2.88 \pm 0.54$ at 95\% C.L. \cite{Neff}, yielding an upper bound \mbox{$f_{10\mathrm{MeV}} < 0.028$}. If the dark matter is still relativistic at BBN, then achieving the observed relic abundance requires \mbox{$f_{10\mathrm{MeV}} \simeq 5 \times 10^{-7} \gamma_D (3 \mathrm{MeV} / T_{\mathrm{RH}}) g_{*S}^{-1/3}(0.34\trh)$}, as given by Eq.~(\ref{eq:pr/px}). Therefore, the upper bound on $f_{10\mathrm{MeV}}$ translates to a bound on \mbox{$\gamma_D$: \mbox{$\gamma_D \lesssim 5.4 \times 10^4 (\trh/3 \mathrm{MeV})g_{*S}^{1/3}(0.34\trh)$}}.

This upper bound on $\gamma_D$ implies an upper bound on the value of $f$ in Eq. (\ref{eq:foftrh}) that can result in the observed relic abundance We will see in Section \ref{sec:lyalpha} that restrictions from small-scale structure on the parameter space of $\gamma_D$ and $\trh$ provide much stronger bounds. As such, annihilations are still necessary to reduce the dark matter abundance to its required value following reheating without fine-tuning $f$.

\subsection{The Adiabatic Cooling of Dark Matter}
\label{sec:redshift}

Given that the momentum of a particle scales as \mbox{$p\propto a^{-1}$}, a particle born from a decay at a scale factor $\ad$, with a physical velocity $\vd$, has a velocity at some later time given by

\begin{align}
v^2(a,a_D) = \frac{\vd^2}{(1 - \vd^2)\left(\frac{a}{\ad}\right)^2+\vd^2}.
\end{align}
The average velocity over all the dark matter particles at any given time is then

\begin{align}\label{eq:vavg}
\langle v^2(a) \rangle = \left[\int_{1}^{a}v^2(a,\ad)\frac{d\hat{n}_\chi}{d\ad}d\ad\right] \left[\int_{1}^{a}\frac{d\hat{n}_\chi}{d\ad}d\ad\right]^{-1},
\end{align}
which can be evaluated in the same manner as Eq. (\ref{eq:avgenergy}).

\begin{figure}
\centering\includegraphics[width=3.4in]{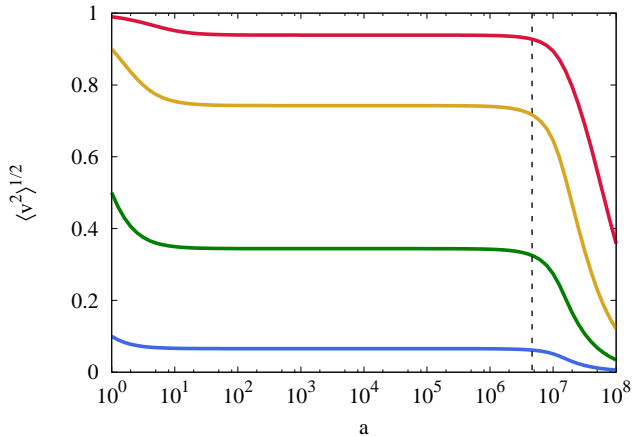}
\caption{The average velocity of the dark matter particles as a function of scale factor $a$ throughout the EMDE for $v_D =$ 0.1, 0.5, 0.9, 0.99 (bottom to top). Reheating is marked by the thin vertical dashed line.}
\label{vavg}
\end{figure}

Figure \ref{vavg} shows the evolution of the average dark matter particle velocity throughout the EMDE until just after reheating; the various curves in Fig. \ref{vavg} represent different values of $\vd$.  We can  see that the average velocity is initially dominated by the few particles born immediately with the imparted velocity.  The velocity of these particles begins to redshift away, pulling the average down, until a steady state is reached between the redshifting of the velocity of existing particles and the creation of new, hot particles.

In the regime where the average velocity has reached a constant value - deep into the EMDE and well before reheating - we can simplify our calculation of the average velocity even further and analytically solve the integrals of Eq. (\ref{eq:vavg}). During the EMDE, the energy density of the Universe is dominated by the scalar field, and our expression in Eq. (\ref{eq:dnda}) becomes
\begin{align}\label{eq:rhophiapprox}
\frac{\rho_\phi}{H_D}a_D^2 \simeq\frac{\rho_{\phi,i}a_D^{-3}}{H_i a_D^{-3/2}}a_D^2 = \frac{\rho_{\phi,i}}{H_i}\sqrt{a_D}.
\end{align}

Rewriting the expression for $v^2$ to make its dependence on the integration variable, $a_D$, more apparent, we have
\begin{align}\label{}
v^2(a,a_D) = \frac{\ad^2}{\left(\frac{a}{X}\right)^2+\ad^2},
\end{align}
where $X \equiv \B$. And so, deep in the EMDE, our expression for the average velocity, given by Eq. (\ref{eq:vavg}), takes the form
\begin{align}\label{}
\langle v^2(a) \rangle = \left[\int_{1}^{a}\frac{a_D^{5/2}}{\left(\frac{a}{X}\right)^2+ a_D^2}d\ad\right]\left[\int_{1}^{a}\sqrt{a_D}d\ad\right]^{-1}.
\end{align}
The solution to the integral in the numerator is given by
\begin{widetext}
\begin{align}\label{}
\int_{1}^{a}\frac{a_D^{5/2}d\ad}{\left(\frac{a}{X}\right)^2+ a_D^2} \ =  \ \ & \frac{2}{3}(a^{3/2} -1) \ \ + \ \ \left(\frac{a}{2X}\right)^{3/2} \ln\left(\frac{1+\sqrt{2X}+X}{1-\sqrt{2X}+X} \ \ \frac{a-\sqrt{2a X}+X}{a+\sqrt{2a X}+X}\right) \\  \nonumber
& + \ \  2\left(\frac{a}{2X}\right)^{3/2}\left[\tan^{\textrm{-}1}(1-\sqrt{2X}) - \tan^{\textrm{-} 1}(1+\sqrt{2X})- \tan^{\textrm{-}1}\left(1-\sqrt{\frac{2X}{a}}\right) + \tan^{\textrm{-}1}\left(1+\sqrt{\frac{2X}{a}}\right)\right].
\end{align}
\end{widetext}
The solution to the integral in the denominator is simply
\begin{align}\label{}
\int_{1}^{a}\sqrt{a_D}d\ad = \frac{2}{3}(a^{3/2} -1).
\end{align}
Long after the decays have started ($a \gg 1$), both integrals scale as $a^{3/2}$ and $\langle v^2(a) \rangle$ is constant until just prior to reheating, at which point our approximation in Eq. (\ref{eq:rhophiapprox}) is no longer valid.

The steady state between the cooling of old particles and the creation of new, hot ones is maintained until just before reheating, and the average dark matter velocity at reheating is not reduced significantly from the velocity imparted at the scalar's decay. Relativistic-born dark matter, $\vd = 0.99$, is still considerably relativistic at reheating, $\sqrt{\langle v^2 \rangle} \simeq 0.93$. At reheating, the average dark matter particle is nonrelativistic only if $v_D$ is already largely nonrelativistic: $\sqrt{\langle v^2 \rangle} \lesssim 0.01$ requires $v_D < 0.017$.

If the comoving size of the horizon at reheating, $k_{\mathrm{RH}}^{-1} = (a_{\mathrm{RH}} H_{\mathrm{RH}})^{-1}$, is smaller than the dark matter free-streaming horizon, $k_{\mathrm{fs}}^{-1}$, then the random drift of dark matter particles will erase the growth of density perturbations that occurred during the EMDE. Reference \cite{Adrienne} found that $k_{\mathrm{RH}} / k_{\mathrm{fs}} < 1$ required the dark matter velocity at reheating to be $v_{\mathrm{RH}} \lesssim 0.06$. Preserving enhanced structure growth requires an even smaller average velocity. Achieving such a small average velocity at reheating would require the dark matter particles to be born with a similarly small velocity.

\section{Dark Matter Distribution Function}
\label{sec:dmdf}

Although the average dark matter particle may have too large a velocity to participate in enhanced structure formation, we would like to investigate what fraction of the dark matter population has a sufficiently low velocity to do so. Instead of considering the average particle velocity at reheating, we will consider what fraction of the dark matter has a velocity at reheating that is less than a percent of the speed of light. To this end, we begin by deriving the dark matter distribution function. At a given time, $a$, the fraction of dark matter with velocities below a particular threshold equals the fraction of dark matter born before the correspondingly required ``birth time,'' $a_D$.

From the fact that $v/\sqrt{1-v^2} \propto a^{-1}$, we know that in order for a particle born with velocity $\vd$ to have a velocity less than $v_{\mathrm{RH}}$ at reheating, that particle must have been born from decay at a scale factor,
\begin{align}\label{eq:acold}
a_D < \frac{v_{\mathrm{RH}}}{v_D}\sqrt{\frac{1-\vd^2}{1-v_{\mathrm{RH}}^2}}a_{\mathrm{RH}}.
\end{align}
Obtaining the distribution function in birth times of the dark matter particles, $f(a_D)$, will then allow us to compute the fraction of dark matter born before this time.

\begin{figure}
\centering\includegraphics[width=3.4in]{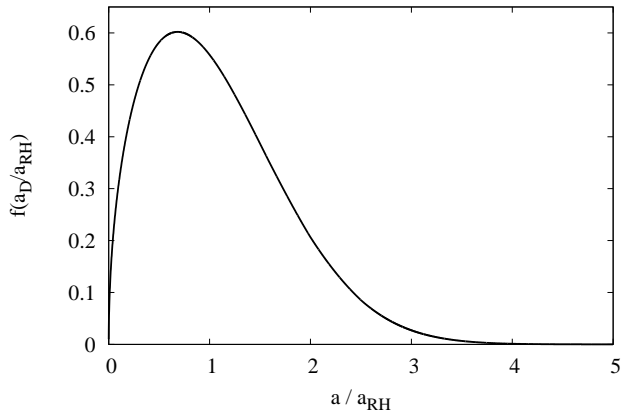}
\caption{The birth time distribution function of dark matter for our model. The peak production of dark matter occurs just prior to reheating and $f(a_D / a_{\mathrm{RH}})$ is maximized at \mbox{$a_D \simeq 0.68 a_{\mathrm{RH}}$.}}
\label{faD}
\end{figure}

The fraction, $\varepsilon$, of dark matter particles born within a particular interval of scale factor, $a_{D,1} > a_D > a_{D,2}$, is given by
\begin{align}\label{}
\varepsilon_{a_{D,12}} = \int_{a_{D,1}}^{a_{D,2}}f(a_D)da_D.
\end{align}
This fraction can also be directly computed by
\begin{align}\label{}
\varepsilon_{a_{D,12}} = \frac{\int_{a_{D,1}}^{a_{D,2}} d\hat{n}_\chi}{\int_{1}^{\infty}d\hat{n}_\chi} = \frac{\int_{a_{D,1}}^{a_{D,2}} \frac{d\hat{n}_\chi}{da_D}da_D}{\int_{1}^{\infty}\frac{d\hat{n}_\chi}{da_D}da_D}.
\end{align}
Equating the two expressions for $\varepsilon_{a_{D,12}}$ and considering small intervals in the scale factor, we can derive an expression for the distribution function:
\begin{align}\label{eq:fad}
\frac{\int_{a_{D,1}}^{a_{D,2}} d\hat{n}_\chi}{\int_{1}^{\infty}d\hat{n}_\chi} &= \int_{a_{D,1}}^{a_{D,2}}f(a_D)da_D \\ \nonumber
& \simeq f(a_D)\Delta a_D,\\ \nonumber
f(a_D) & \simeq \frac{\int_{a_{D,1}}^{a_{D,2}} d\hat{n}_\chi}{\Delta a_D \int_{1}^{\infty}d\hat{n}_\chi}.
\end{align}
Numerically evaluating Eq. (\ref{eq:fad}), we obtain the distribution function of dark matter birth times seen plotted in Fig. \ref{faD} as $f(a_D / \arh) = \arh f(a_D)$. From Fig. \ref{faD}, one can see that approximately half of the dark matter is born after reheating, $a/\arh >1$.

\begin{figure}
\centering\includegraphics[width=3.4in]{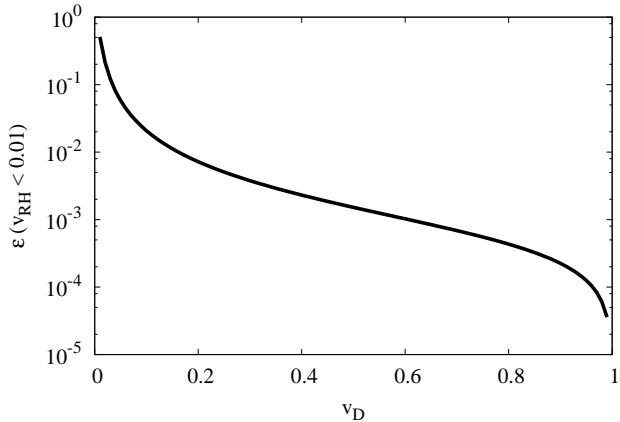}
\caption{Fraction of dark matter whose velocity at reheating is $v_{\mathrm{RH}} < 0.01$ as a function of the assumed velocity imparted to the dark matter at its production.}
\label{coldfracPlot}
\end{figure}

From Eq. (\ref{eq:acold}) one can find that, even for dark matter particles imparted with a velocity only half of the speed of light, only those born before $a_D \lesssim 0.017 a_{\mathrm{RH}}$ will have a velocity at reheating $v_{\mathrm{RH}} < 0.01$.  Integrating our distribution function over this interval in $a_D$, we find the fraction of dark matter born before this time to be approximately 0.15\%. Figure \ref{coldfracPlot} shows the fraction of dark matter that has a velocity below $v=0.01$ at reheating as a function of the given value of $\vd$. Even for dark matter born at only one tenth of the speed of light, only $\sim 2\%$ of the dark matter has the required $v_{\mathrm{RH}} < 0.01$.

One of the intriguing consequences of an EMDE is that density perturbations in matter grow linearly with scale factor during an EMDE, which is faster than the logarithmic growth expected during the typically assumed radiation-dominated epoch. We now examine what fraction of the dark matter is able to retain an appreciable perturbation enhancement from this linear growth. Since density perturbations grow linearly during the EMDE, a mode that enters the horizon at a scale factor of $0.1 a_{\mathrm{RH}}$ will grow by a factor of $\sim$10 during the EMDE, which we will consider ``appreciable''. The comoving wavelength of such a mode is given by the horizon size at this time:
\begin{align}\label{eq:hor}
\lambda = \left.\lambda_{\mathrm{hor}}\right|_{a_{\mathrm{RH}}/10} = \frac{1}{\frac{a_{\mathrm{RH}}}{10}\ H\! \left(\frac{a_{\mathrm{RH}}}{10}\right)}.
\end{align}
Any modes that enter the horizon prior to $0.1 a_{\mathrm{RH}}$ will experience even more growth. The comoving free-streaming length of a dark matter particle born at $a_D$ is given by
\begin{align}\label{eq:fs}
\lambda_{\mathrm{fs}} = \int_{a_D}^{a_0} v(a) \frac{da}{a^2 H(a)},
\end{align}
where $a_0$ is the value of the scale factor today. Similar to our approach in the previous evaluation, there is a value of $a_D$ for which the dark matter free-streaming length is less than the horizon size at $0.1 a_{\mathrm{RH}}$ ($\lambda_{\mathrm{fs}} < \lambda_{\mathrm{hor}} |_{a_{\mathrm{RH}}/10}$).

\begin{figure}
\centering\includegraphics[width=3.4in]{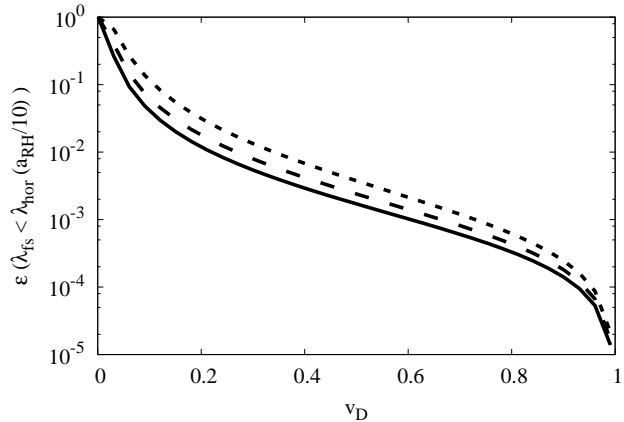}
\caption{Fraction of dark matter whose free-streaming length is smaller than the scale of the perturbation mode that experiences a factor of 10 in growth during the EMDE as a function of the assumed dark matter velocity at production. The different lines represent values of $f = 10^{-7}$ \ (solid), $10^{-5}$ \ (dashed), and $10^{-3}$ (dotted), or equivalently, $T_{\mathrm{RH}} \simeq 3\gamma_D$ MeV, $T_{\mathrm{RH}} \simeq 0.03\gamma_D$ MeV, and $T_{\mathrm{RH}} \simeq 0.0003\gamma_D$ MeV, respectively, in order to produce the observed relic abundance of dark matter without annihilations.}
\label{slowfrac}
\end{figure}

\begin{figure}
\centering\includegraphics[width=3.4in]{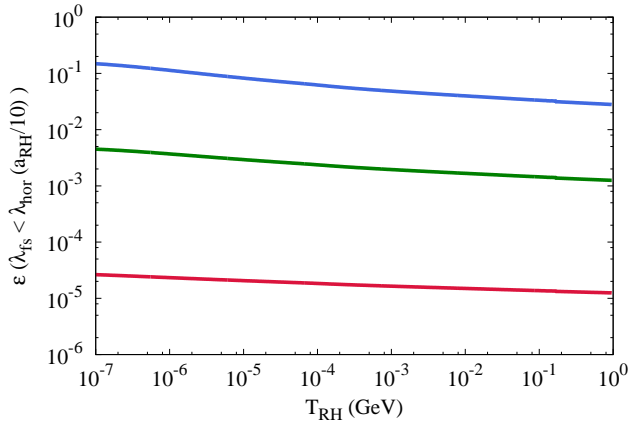}
\caption{Fraction of dark matter whose free-streaming length is smaller than the scale of the perturbation mode that experiences a factor of 10 in growth during the EMDE as a function of the reheat temperature. The different lines represent, from top to bottom, values of $v_D = 0.1$, $0.5$, and $0.99$.}
\label{slowfractrh}
\end{figure}

The resulting fraction of dark matter that is born before this time, and thus that preserves a factor of 10 or more growth in perturbation amplitude, is shown in Fig. \ref{slowfrac}. The integral in Eq. (\ref{eq:fs}) can be broken into three separate contributing integrals, representing the scalar-, radiation-, and matter-dominated eras (because the dark matter free-streaming length does not change significantly after matter-radiation equality, we neglect dark energy). The contribution coming from the radiation-dominated era is dependent on the duration of the era, which, in our formalism, is set by the relative abundance of dark matter and radiation following reheating, and this is directly related to the reheat temperature. The dependence on $T_{\mathrm{RH}}$, however, is only logarithmic and large variations in $T_{\mathrm{RH}}$ do not result in significant changes in the resulting fraction. Due to the interdependency discussed in Section \ref{sec:1} between the parameters $f$, $T_{\mathrm{RH}}$, and $v_D$ required to obtain the appropriate dark matter abundance, we plot the fraction of dark matter able to preserve enhanced structure growth both as a function of $v_D$ for various $f$ in Fig. \ref{slowfrac} and as a function of $T_{\mathrm{RH}}$ for various $v_D$ in Fig. \ref{slowfractrh}. Figure \ref{slowfractrh} shows how insensitive the fractional component of dark matter that experiences enhanced structure growth is to the reheat temperature.

Unfortunately, for dark matter particles born relativistically throughout an EMDE, the redshifting of their momentum is not enough to allow an appreciable fraction of the particles to participate in enhanced structure formation. Studies of mixed dark matter, in which there are both cold and warm dark matter components, show that the small-scale matter power spectrum is suppressed by $99\%$ even when up to half of the dark matter is cold \cite{mixed}. Therefore, the fraction of dark matter that is cold enough to benefit from the growth of perturbations during the EMDE is far too small for these structures to form.

\section{Lyman-Alpha and MW Satellite Constraints}
\label{sec:lyalpha}

We have shown that the redshifting of the momentum of dark matter particles prior to reheating does not cool the dark matter enough to preserve the enhanced structure growth on scales that enter the horizon during the EMDE ($\lambda \lesssim 30$ pc for $T_{\mathrm{RH}} > 3$ MeV). In this section we consider if the dark matter is too hot, i.e. if its free-streaming length large enough to prevent the formation of the smallest observed structures. Analysis of Lyman-$\alpha$ data can be used to probe the matter power spectrum on small scales, $0.5\mathrm{Mpc}/h < \lambda < 20 \mathrm{Mpc}/h$ \cite{lyman1,lyman2,lyman3}, or $ 12.6 h/\mathrm{Mpc} > k > 0.06 h/\mathrm{Mpc}$,  and we compare the degree of gravitational clustering at these scales in our model to that of the traditional model of cold dark matter. The existence of MW satellite galaxies provides another probe of small-scale structure formation.  Suppression of the power spectrum leads to an underabundance of small structures, and the known abundance of substructures in the vicinity of the MW provides a bound on the allowed suppression \cite{Milkway}.

\subsection{Free-Streaming Length}

We begin by calculating the physical streaming length today of a particle born at reheating:
\begin{align}\label{eq:fsrh}
\lambda_{\mathrm{fs},0}^{\mathrm{phys}} = a_0 \int_{a_{\mathrm{RH}}}^{a_0} v(a) \frac{da}{a^2 H(a)}.
\end{align}
The choice of $a_D = a_{\mathrm{RH}}$ is beneficial in that the fraction of dark matter born before reheating, found by our previous analysis of the distribution of birth times, is $0.51$, and we can say that approximately half of the dark matter will have a free-streaming length above or below our calculated value. Another benefit of this choice is that our calculations of the free-streaming length are made simpler by neglecting the contribution to the free-streaming length coming from the EMDE.

Calculating the free-streaming length using Eq. (\ref{eq:fsrh}) shows that $> 75\%$ of this distance is covered after the dark matter particle has become nonrelativistic, (\mbox{$\gamma \leq 1.01$}). For the highly relativistic initial velocities we would like to consider, the dark matter particles remain relativistic well after reheating, and are still relativistic after changes in the number of relativistic degrees of freedom have ceased. Beginning the integral in Eq. (\ref{eq:fsrh}) at $a_*$, the value of the scale factor after which $g_*$ remains constant, captures most ($\gtrsim 90\%$) of the free-streaming length and illuminates the important features of this scenario by allowing us to assume that $H \propto a^{-2}$ during radiation domination.

\begin{figure}
\centering\includegraphics[width=3.4in]{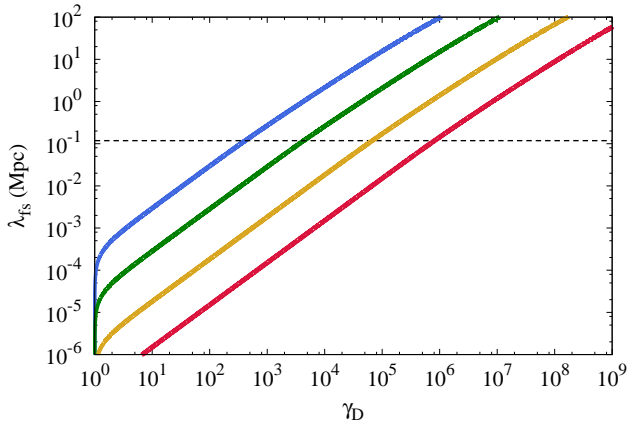}
\caption{The free-streaming length of the dark matter as calculated by Eq. (\ref{eq:lfsphys0}) as a function of the Lorentz factor at decay, $\gamma_D$, for (left to right) $T_{\mathrm{RH}} = (3, 30, 300, 3000)$ MeV.  The horizontal dashed line marks $\lambda_{\mathrm{fs},0}^{\mathrm{phys}} \gtrsim 0.08 \ \mathrm{Mpc}/h$, approximately the lower limit to scales probed by Lyman-$\alpha$ observations.}
\label{lfsPlot}
\end{figure}

We begin by breaking the integral into two separate contributing integrals, representing the radiation- and matter-dominated eras (again we neglect dark energy) and introducing the variable $Y \equiv (\B a_D)^2$. The free-streaming length is then
\begin{widetext}
\begin{align}\label{}
\lambda_{\mathrm{fs}} = \int_{a_*}^{a_0} \sqrt{\frac{Y}{Y+a^2}} \frac{da}{a^2 H(a)} & \simeq \frac{1}{H_{*}a_{*}^2}\int_{a_*}^{a_{eq}} \sqrt{\frac{Y}{Y+a^2}}\ \ da + \frac{1}{ H_{\mathrm{eq}} \aeq^{3/2}} \int_{\aeq}^{a_0} \sqrt{\frac{Y}{Y+a^2}} \frac{da}{\sqrt{a}} \nonumber \\
& = \frac{\sqrt{Y}}{H_*a_*^2} \ln\left({\frac{\aeq+\sqrt{Y + \aeq^2}}{a_*+\sqrt{Y +a_*^2}}}\right) + \frac{2 \sqrt{i\sqrt{Y}}}{H_{\mathrm{eq}} \aeq^{3/2}} \left.F\left(i \sinh^{-1}\sqrt{ \frac{i\sqrt{Y}}{a}},-1\right)\right|^{a_0}_{\aeq},
\end{align}
where $F(\phi,m)$ is the elliptic integral of the first kind. By choosing $a_D = a_{\mathrm{RH}}$ (contained in the variable $Y$), the first term in the above expression can be simplified under the assumption that $\aeq \gg \arh$, which is reasonable considering that matter-radiation equality occurs at a temperature of $T_{\mathrm{eq}} \simeq 0.8$ eV and we require $T_{\mathrm{RH}}>3$ MeV. The expression for the contribution to the physical free-streaming length today coming from the radiation-dominated era then simplifies to
\begin{align}\label{eq:lfsphysRD}
\lambda_{\mathrm{fs},0}^{\mathrm{RD}} & \simeq \frac{\B \arh a_0}{H_* a_*^2} \ln\left(\frac{\aeq}{a_*}\frac{2}{1+\sqrt{\left(\B \frac{\arh}{a_*}\right)^2+1}}\right); \nonumber \\
& = \left(4.66\times 10^{11}\mathrm{pc}\right) \B\frac{\arh}{a_0}\left[\ln\left(2\frac{T_*}{T_{\mathrm{eq}}}\right)- \ln\left(1+\sqrt{\left(\B\frac{\arh}{a_0}\right)^2\left(\frac{T_*}{T_0}\right)^2+1}\right)\right],
\end{align}
where we have used the fact that $g_*$ remains constant after $T_* = 2\times10^{-5}$ GeV to set $a_*T_*=\aeq T_{\mathrm{eq}}=a_0T_0$. An important feature of this calculation is that the parameters of our model, the dark matter velocity at its production and the reheat temperature, only enter into this expression through the combination
\begin{align}\label{eq:mu}
\mu \equiv \frac{\B\arh}{a_0} = \B \frac{T_0 g_{*S,0}^{1/3}}{3\left[Tg_{*S}^{1/3}\right]_{T=0.34\trh}},
\end{align}
where $g_{*S}$ is the number of relativistic degrees of freedom in the entropy density and again we assume entropy is conserved after $a=3\arh$. Expressed in terms of the variable $\mu$, the physical free-streaming length calculated from the contribution from both the radiation- and matter-dominated eras is
\begin{align}\label{eq:lfsphys0}
\lambda_{\mathrm{fs},0}^{\mathrm{phys}}  = & \left(4.66\times 10^{5}\mathrm{Mpc}\right) \mu \left[\ln\left(2\frac{T_*}{T_{\mathrm{eq}}}\right)- \ln\left(1+\sqrt{\mu^2 \left(\frac{T_*}{T_0}\right)^2 + 1}\right)\right] \nonumber \\
& + \left(4.66\times 10^{5}\mathrm{Mpc}\right)\sqrt{\frac{T_0}{T_{\mathrm{eq}}}} \sqrt{2 i \mu} \left[F\left(i\sinh^{-1}\sqrt{i\mu}\right) - F\left(i\sinh^{-1}\sqrt{i\mu\frac{T_{\mathrm{eq}}}{T_0}}\right)\right].
\end{align}
\end{widetext}
The above equation gives the scale at which the power spectrum of our model begins to differ from that of the standard $\Lambda$CDM power spectrum. Figure \ref{lfsPlot} shows the free-streaming length calculated by Eq. (\ref{eq:lfsphys0}) as a function of the Lorentz factor at decay, $\gamma_D$, for different values of the reheat temperature.  We define $k_{\mathrm{fs}} = (\lambda_{\mathrm{fs},0}^{\mathrm{phys}})^{-1}$, and above the horizontal dashed line, the free-streaming length of the dark matter reaches scales probable by the Lyman-$\alpha$ forest: $k \lesssim 12.6 \ h/\mathrm{Mpc}, \lambda_{\mathrm{fs},0}^{\mathrm{phys}} \gtrsim 0.08 \ \mathrm{Mpc}/h$. Since the free-streaming lengths of our model enter the observable regime, we consider a more precise determination of the effects of the dark matter free-streaming length in the next section.

\subsection{Transfer Function}
\label{sec:Tk}

We use the Cosmic Linear Anisotropy Solving System (CLASS) \cite{CLASS} to obtain the dark matter transfer function,

\begin{align}
T^2(k) \equiv \frac{P_{\mathrm{nCDM}}(k)}{P_{\mathrm{CDM}}(k)},
\end{align}
which describes suppression of structure due to non-cold dark matter (nCDM) compared to that of the standard CDM scenario; $P_{\mathrm{nCDM}}(k)$ and $P_{\mathrm{CDM}}(k)$ are the matter power spectra in each respective case.

\begin{figure}
\centering\includegraphics[width=3.4in]{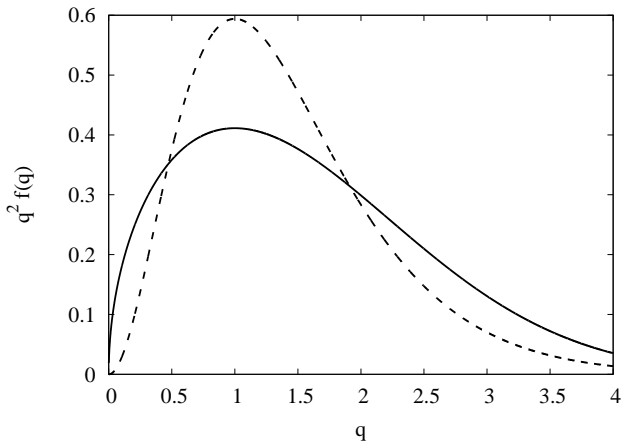}
\caption{The distribution function of dark matter for our model (solid) and a Fermi-Dirac distribution (dashed) for comparison.}
\label{psdf}
\end{figure}

Acquiring the power spectrum for our scenario requires us to determine the momentum distribution function of our dark matter model. Shortly after reheating ($a \sim 3 a_{\mathrm{RH}}$), the scalar field has decayed almost entirely, and essentially no new dark matter particles are being produced. After this point, the distribution of the comoving momenta of the dark matter particles does not change. We scale the comoving momenta of the dark matter particles by the comoving momentum of a particle born at the scale factor that maximizes $f(a_D)$, $a_{\mathrm{max}} = 0.68 a_{\mathrm{RH}}$, and express our distribution function in terms of
\begin{align}\label{}
q \equiv \frac{a p}{a_\mathrm{max} p_D} = \frac{a_D}{a_\mathrm{max}},
\end{align}
where $p_D$ is the physical momentum of a particle with velocity $v_D$. Since we assume that all dark matter particles are produced with the same velocity, the distribution in momentum for particles in our scenario can be entirely determined from the distribution in the particles' scale factor at production, which we have already determined. The two distribution functions can be related through
\begin{align}\label{}
4\pi q^2 f(q) = f(a_D) \frac{da_D}{dq} = 0.68 \ f\left(\frac{a_D}{\arh}\right).
\end{align}
This distribution function is shown in Fig. \ref{psdf}, and we also show for comparison the Fermi-Dirac distribution that is maximized at $q=1$. We can see that, compared to the thermal case, we have a broader distribution function.

\begin{figure}
\centering\includegraphics[width=3.4in]{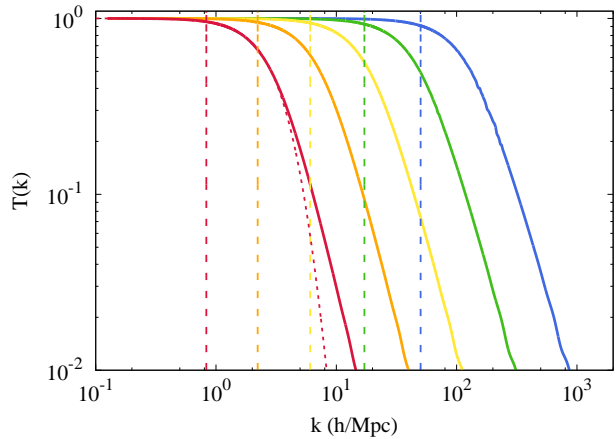}
\caption{The transfer function for several values of the dark matter velocity at production, $\vd$.  Right to left (cool to warm colors), the solid lines represent dark matter produced with increasing Lorentz factor $\gamma_D = 100, 300, 900, 2700, 8100$, respectively. In all cases $T_{\mathrm{RH}} = 3$ MeV. Vertical dashed lines represent the scale of the free-streaming horizon calculated using Eq. (\ref{eq:lfsphys0}). The red dotted line represents the typical transfer function for dark matter with a thermal distribution ($\beta_\mathrm{WDM} = 2.24$ and $\gamma_\mathrm{WDM} = -4.46$) with a cutoff parameter $\alpha_\mathrm{WDM} \simeq 0.16$ Mpc$/h$ in order to match the same half-mode scale $\khm$ as our far left curve with $\alpha = 0.31$ Mpc$/h$.}
\label{tk}
\end{figure}

\begin{figure}
\centering\includegraphics[width=3.4in]{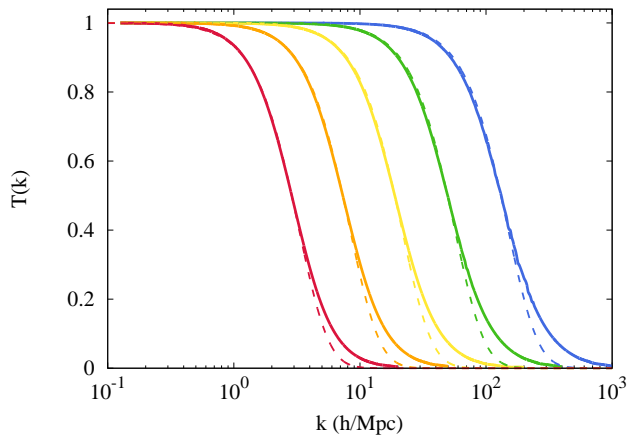}
\caption{The solid lines show the same transfer functions shown in Fig. \ref{tk}. Dashed lines show the thermal WDM transfer functions with matched half-mode scales. Due to the difference in the shape parameters $\beta$ and $\gamma$ in the fitting form of Eq. (\ref{eq:tkfit}) between the two models, matching the half-mode scales requires the cutoff parameters, $\alpha$, to be related by approximately $\alpha \simeq  2\alpha_\mathrm{WDM}$}
\label{tkhalfmatch}
\end{figure}

With our distribution function $f(q)$, we are able to use CLASS to obtain transfer functions for any combination of the velocity imparted to the dark matter and the reheat temperature by also providing the present-day physical momentum of a dark matter particle with $q=1$:
\begin{align}\label{eq:p0}
p_0 = \frac{a_\mathrm{max}p_D}{a_0} \propto \frac{\arh \B}{a_0}.
\end{align}
Again we find that, just as in our calculations of the free-streaming length, the relevant combination of parameters is $\mu = \B \arh / a_0$.  In Figs. \ref{tk} and \ref{tkhalfmatch} we show the transfer functions for dark matter produced at different velocities, but in scenarios with the same reheat temperature, 3 MeV. As expected, dark matter particles born at greater velocities result in the suppression of larger scales (smaller $k$). The vertical dashed lines in Fig. \ref{tk} mark the free-streaming horizon $k_{\mathrm{fs}} = (\lambda_{\mathrm{fs},0}^{\mathrm{phys}})^{-1}$ given by Eq. (\ref{eq:lfsphys0}) for each of the different velocities at production, confirming that it is the scale at which our model begins to show deviation, $T(k) \simeq 0.95$, from the CDM scenario.

\begin{figure}[b]
\centering\includegraphics[width=3.4in]{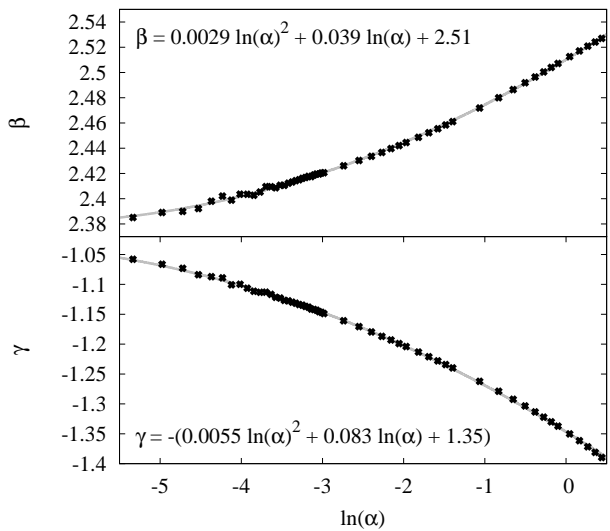}
\caption{The fitted values for the parameters $\beta$ and $\gamma$ for transfer functions whose cutoff parameters, $\alpha$, span from 0.005 to 1.5 Mpc$/h$. In the fitting functions for $\beta$ and $\gamma$, $\alpha$ has units of Mpc$/h$.}
\label{betagamma}
\end{figure}

Transfer functions in nCDM models, such as this, can be well described by a fitting formula \cite{ApprovEd}:
\begin{align}\label{eq:tkfit}
T(k) = [1+(\alpha k)^\beta]^\gamma.
\end{align}
Using Lyman-$\alpha$ data, the fitting parameters $\alpha, \beta$, and $\gamma$ can be constrained \cite{ApprovEd,ApprovEd2}, and the parameters of our model, $v_D$ and $T_{\mathrm{RH}}$, can be constrained as well. The typical scale of the suppression is set by $\alpha$, whereas the general shape is determined by $\beta$ and $\gamma$. When fitting our transfer function at values $T(k) > 0.01$, the overall shape of our transfer function varies little across wide ranges of our parameter space, and $\beta$ and $\gamma$ can be expressed as functions of $\alpha$, as seen in Fig. \ref{betagamma}. The cutoff parameter $\alpha$ is then our only free parameter, and it can be robustly constrained using Lyman-$\alpha$ data.

We find typical values of $\beta$ and $\gamma$ for our model to be approximately 2.4 and $-1.1$ respectively, for $\alpha$ near the constrainable regime. These values are noticeably different from those that describe the thermal warm dark matter (WDM) transfer function, $\beta_\mathrm{WDM} = 2.24$ and \mbox{$\gamma_\mathrm{WDM} = -4.46$}. If we compare our transfer functions to those of WDM with the same value of the half-mode scale\footnote{We follow the convention of Ref. \cite{schneider} and define the half-mode scale via $T(\khm)=0.5$, noting that this convention is different from the half-mode scale, $k_{1/2}$, defined in Refs. \cite{ApprovEd} and \cite{ApprovEd2}, for which $T^2(k_{1/2})=0.5$.} $\khm$, we can see in Figs. \ref{tk} and \ref{tkhalfmatch} that the transfer functions in the two models are quite similar. However, due to the difference in the shape parameters of the transfer function fit between the two models, matching their half-mode scales requires the cutoff parameter in the WDM transfer function to be roughly a factor of $2$ smaller than that in the corresponding nonthermal transfer function, $\alpha \simeq  2\alpha_\mathrm{WDM}$. The cutoff in the transfer function of our model is not as sharp as that of WDM, but they only begin to differ significantly at scales at which the power in the nCDM model is already greatly suppressed, $T(k) \lesssim 0.1$.

Reference \cite{ApprovEd2} provided marginalized bounds on all three fitting parameters in Eq. (\ref{eq:tkfit}). The predictable shape of our transfer function determines the values of $\beta$ and $\gamma$ as a function of $\alpha$, as shown in Fig. \ref{betagamma}, and allows us to obtain a bound on the remaining free parameter, the scale of the cutoff: $\alpha < 0.011 \mathrm{Mpc}/h$ (68\% C.L.) and $\alpha < 0.026 \mathrm{Mpc}/h$ (95\% C.L.). Following Ref. \cite{ApprovEd2}, these limits have been obtained by performing a comprehensive Markov hain Monte Carlo (MCMC) analysis of the full parameter space affecting the one-dimensional flux power spectrum, which is the Lyman-$\alpha$ forest physical observable, with a data set consisting of the high-resolution and high-redshift ($4.2 < z < 5.4$) quasar samples from MIKE and HIRES spectrographs \cite{riccardo1}. The flux power spectra to be compared against observations are estimated by interpolating in the multidimensional space defined by the sparse grid of precomputed hydrodynamic simulations described in Ref. \cite{ApprovEd2}. Whenever some of the parameters assume values not enclosed by the template of simulations, the corresponding values of the power spectra are linearly extrapolated.

As in the reference analysis from Ref. \cite{ApprovEd2} (but see also, e.g., Refs. \cite{riccardo2} and \cite{riccardo3}), the other cosmological and astrophysical parameters impacting our likelihood are treated as nuisance parameters to marginalize over. We adopt conservative flat priors on both $\sigma_8$, i.e. the normalization of the linear matter power spectrum, and $n_{\mathrm{eff}}$, i.e. the slope of the matter power spectrum at the scale of the Lyman-$\alpha$ forest ($k \sim 1 h/$Mpc), in the intervals $[0.5,1.5]$ and $[-2.6,-2.0]$, respectively; and on the instantaneous reionization redshift $z_\mathrm{reio}$ (in the range $[7,15]$). Concerning the astrophysical parameters, we model the redshift evolution of the temperature of the intergalactic medium as a power law, imposing flat priors on both its amplitude and tilt (once again, see Ref. \cite{ApprovEd2} for further details). Finally, we adopt conservative Gaussian priors on the mean Lyman-$\alpha$ forest fluxes $\langle F(z)\rangle $, with standard deviation $\sigma = 0.04$ \cite{lyman3}, and a flat prior on $f_\mathrm{UV}$, which is an effective parameter accounting for spatial ultraviolet fluctuations in the ionizing background.

Last but not least, we adopt a flat prior on $\alpha$ in the interval [0, 0.1] Mpc$/h$, while the parameters $\beta$ and $\gamma$ are derived analytically, per each MCMC step, according to the expressions reported in Fig. \ref{betagamma}. For further details on the data set, simulations, and methods that we have used, we address the reader to any of the aforementioned references \cite{ApprovEd2,riccardo2,riccardo3}.

\begin{figure}
\centering\includegraphics[width=3.4in]{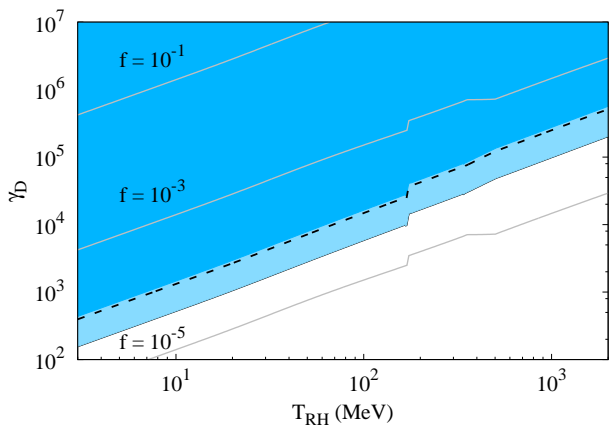}
\caption{Limits on the Lorentz factor $\gamma_D$ and the reheat temperature. The shaded regions correspond to the 1$\sigma$ and 2$\sigma$ bounds, $\alpha = 0.011 \mathrm{Mpc}/h$ and $0.026 \mathrm{Mpc}/h$, respectively. The thick dashed line represents $k_{\mathrm{fs}} = (\lambda_{\mathrm{fs}})^{-1} = 12.6 h/\mathrm{Mpc}$ as calculated by Eq. (\ref{eq:lfsphys0}). The discontinuity at $T_{\mathrm{RH}} \simeq 170$MeV occurs due to the sudden change in $g_{*}$ during the QCD phase transition. Thin solid lines show the contours of $f$ required to obtain the observed dark matter abundance in the absence of annihilations. The thick solid line shows the bound on $\gamma_D$ as a function of $\trh$ imposed by BBN, derived in Section \ref{sec:abundance}.}
\label{gtrhbounds}
\end{figure}

For comparison, just as matching the half-mode scale of the thermal WDM transfer function with that of our nonthermal model requires $\alpha \simeq 2\alpha_\mathrm{WDM}$, the $\alpha_\mathrm{WDM}$ value of a $3$ keV WDM particle, $\alpha_\mathrm{WDM} \simeq 0.015$ Mpc$/h$ \cite{ApprovEd2}, is approximately a factor of $2$ smaller than that of our 95\% C.L. bound on $\alpha$. Our limits on $\gamma_D$ and $T_{\mathrm{RH}}$ corresponding to our 68\% and 95\% C.L. bounds on $\alpha$ are shown in Fig. \ref{gtrhbounds}. Scenarios in which the dark matter is born at too high of a velocity (large $\gamma_D$) or in which the radiation-dominated era is too short (low $T_{\mathrm{RH}}$) are part of the excluded parameter space for our relativistic nonthermal dark matter model. The thin grey lines represent contours of constant $f$, the fraction of the scalar's energy imparted to the dark matter particle that is required to obtain the correct relic abundance for a given reheat temperature without dark matter annihilations. We can see that, in the absence of annihilations, the allowed values of $\gamma_D$ are not large enough for the required value of $f$ to be of order unity.

\begin{figure}
\centering\includegraphics[width=3.4in]{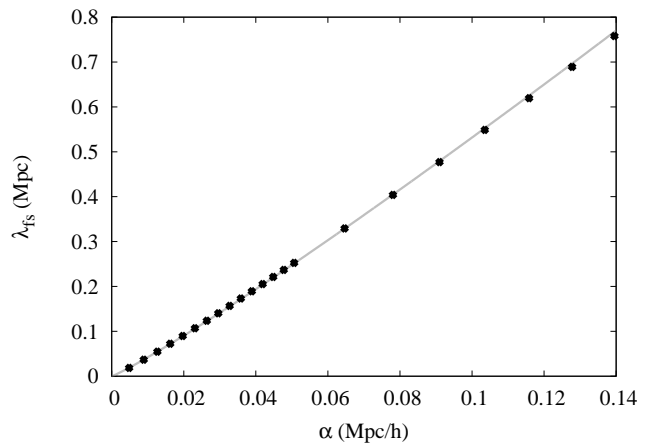}
\caption{A plot of the relationship between the fitting parameter $\alpha$ and the free-streaming length as calculated by Eq. (\ref{eq:lfsphys0}).  Black dots represent points for which we have used our model parameters to calculate the free-streaming length and obtain the transfer function using CLASS.  The grey line shows the fit to the data given by Eq. (\ref{eq:lambdaalpha}). }
\label{lambdaalphaPlot}
\end{figure}

We also show the outline (dashed) of the parameter space in which the free-streaming length, calculated by Eq. (\ref{eq:lfsphys0}), is naively probable by Lyman-$\alpha$ data, i.e. $k_{\mathrm{fs}} < 12.6 h/\mathrm{Mpc}$. As can be seen in Fig. \ref{gtrhbounds}, limiting the free-streaming length provides a bound that is comparable to those obtained from the full consideration of effects to the matter power spectrum; the free-streaming scales of a particle on the boundary of our 68\% and 95\% C.L. regions are $k_{\mathrm{fs}} = 11.4 h/\mathrm{Mpc}$ and $28  h/\mathrm{Mpc}$, respectively. While examining effects on the matter power spectrum leads to more robust bounds within our parameter space, calculations of the free-streaming length are more readily performed. Fortunately, as both the free-streaming length and the dark matter distribution function depend on the same combination of our parameters, $\B \arh / a_0$, there is a simple relationship between the scale of suppression, $\alpha$, and the free-streaming length calculated by Eq. (\ref{eq:lfsphys0}):
\begin{align}\label{eq:lambdaalpha}
\alpha \simeq &  \ 0.177 \left(\frac{\lambda_{\mathrm{fs},0}^{\mathrm{phys}}}{\mathrm{Mpc}}\right)^{0.908}\frac{\mathrm{Mpc}}{h}
\end{align}
We show this relationship in Fig. \ref{lambdaalphaPlot}. In our model, the bounds on $\alpha$ can be easily used to limit the free-streaming length, and thereby the parameters $\gamma_D$ and $T_{\mathrm{RH}}$ on which it depends.

\subsection{Milky-Way Satellites}
\label{sec:MW}

In addition to structures inferred by Lyman-$\alpha$ data, we can also constrain our model using observed structures in the Milky Way. Simulations of thermal warm dark matter provide an indication of how the suppression expected in the matter power spectrum decreases the abundance of collapsed objects. The subhalo mass function in simulations with WDM characterizes this underabundance \cite{Lovell},
\begin{align}\label{eq:mhm}
\left.\frac{dN}{dM}\right|_{\mathrm{WDM}} = \left.\frac{dN}{dM}\right|_{\mathrm{CDM}} \left(1 + \delta \frac{\Mhm}{M} \right)^{-\varepsilon},
\end{align}
where $M$ is the subhalo mass, $\delta = 2.7$ and $\varepsilon = 0.99$, and $\Mhm$ is the mass scale associated with the half-mode scale\footnote{We have verified with the authors of Ref. \cite{Lovell} that they used the same definition of $\khm$ that we have presented here.}:
\begin{align}\label{eq:Mhm}
\Mhm = \frac{4\pi}{3} \Omega_{\mathrm{DM}} \ \rho_\mathrm{crit,0} \left(\frac{\pi}{\khm}\right)^3,
\end{align}
where $\Omega_{\mathrm{DM}}$ is the fraction of the critical density in dark matter.

Knowing the abundance of satellites of our own galaxy, constraints can be placed on the amount of allowed suppression in the subhalo mass function. Using a probabilistic analysis of the MW satellite population and marginalizing over astrophysical uncertainties, Ref. \cite{Milkway} found an upper limit on $\Mhm$ in Eq. (\ref{eq:mhm}), \mbox{$\Mhm < 3.1 \times 10^8 M_\odot$} (95\% C.L.), which implies that the half-mode scale must satisfy $\khm > 36 h/\mathrm{Mpc}$. This bound on $\khm$ can be used to constrain any dark matter model that has a transfer function comparable to WDM, as we have shown ours to be in Fig. \ref{tkhalfmatch}. Though the transfer function of our model does differ slightly from that of WDM, the differences occur only when the nCDM power spectrum is already greatly suppressed compared to that of CDM, $T(k) \lesssim 0.1$.  The fitting form to our transfer function, Eq. (\ref{eq:tkfit}), implies that the half-mode scale is given by:
\begin{align}\label{}
\khm = \frac{1}{\alpha}\left[\left(\frac{1}{2}\right)^{1/\gamma}-1\right]^{1/\beta},
\end{align}
and the constraint $\khm > 36 \ h/$Mpc directly translates to $\alpha < 0.026 \ \mathrm{Mpc}/h$, matching our 95\% C.L. bound from Lyman-$\alpha$ constraints.

\section{Conclusion}
\label{sec:end}

The inclusion of a period of effective matter domination between inflation and BBN is an amply motivated alternative to the standard thermal history of the Universe. If dark matter is produced nonthermally during this era, the viable parameter space for the dark matter annihilation cross section widens greatly, as large ranges of production and annihilation efficiencies can combine to result in the correct relic abundance.

Nonstandard thermal histories could potentially have observable consequences. Unlike the typically assumed period of radiation domination following inflation, in which subhorizon density perturbations grow logarithmically, EMDEs provide an era of linear growth. Linear growth would enhance structure formation on scales that enter the horizon during this era, possibly leading to observable effects. However, in the absence of fine-tuning, it is likely that dark matter produced nonthermally will be imparted with relativistic velocities, and its subsequently large free-streaming length will wipe out this enhancement to structure formation.

By investigating the velocity evolution and distribution of dark matter produced nonthermally from the decay of a massive scalar field, we have confirmed that retaining the linear enhancement to structure growth requires the dark matter to be produced largely nonrelativistic. Despite the early creation of many particles, and their loss of momentum due to adiabatic cooling, the continuous creation of new, hot particles prevents the average dark matter velocity from decreasing appreciably during the EMDE. The average particle at reheating is nearly as relativistic as those newly produced from decay. And because a majority of the dark matter is created around reheating, essentially negligible fractions of dark matter particles have velocities low enough to preserve enhanced structure formation.

We next investigated the upper limit on the dark matter velocity required to preserve the structures we observe. Dark matter particles born with relativistic velocities have free-streaming lengths that may also washout observed small-scale structures.  Lyman-$\alpha$ forest data provides the best-known probes of inhomogeneity at small scales, and ensuring that structure formation at these scales is not observably suppressed constrains the parameter space of nonthermal dark matter.

Using the software CLASS, we obtained the matter power spectrum resulting from our model of nonthermal dark matter.  A transfer function was used to compare our spectrum to that of the standard CDM scenario and showed a cutoff in the power at small scales in our nonthermal scenario similar to that due to WDM. We fit the form of our transfer functions using three free parameters, one of which, $\alpha$ describes the scale of the cutoff in the transfer function and the other two describe its overall shape. The shape of our transfer function varies slightly with the cutoff scale and the two parameters describing its shape are well determined by analytic functions of $\alpha$.

We obtained limits on the allowed scale of the cutoff in the transfer function by performing a comprehensive MCMC analysis using Lyman-$\alpha$ observations: $\alpha < 0.011 \mathrm{Mpc}/h$ (68\% C.L.) and $\alpha < 0.026 \mathrm{Mpc}/h$ (95\% C.L.). From this constraint, we were able to place limits on the allowed velocity imparted at scalar decay for a given temperature at reheating, summarized in Fig. \ref{gtrhbounds}. We also found a simple relation between $\alpha$ and the dark matter free-streaming length that allows one to use the limits on $\alpha$ to limit the free-streaming length which can be calculated analytically from $\gamma_D$ and $\trh$.

Observations of the abundance of MW satellite galaxies provide another probe of the small-scale power spectrum. Using the halo-mass function obtained from WDM simulations, limits on the cutoff scale can also be placed on the WDM transfer function by requiring consistency between the decreased abundance of collapsed objects expected in WDM scenarios, compared to CDM, and the abundance of satellites observed orbiting the MW. These constraints are applicable to any model of dark matter with a transfer function comparable to that of WDM. Comparison of the parameter values that fit the transfer function of our model to those that fit WDM naively imply marked differences between the two models; however, matching the transfer functions at the same half-mode scale shows the two models to be remarkably similar, differing significantly only on scales at which the power was already greatly suppressed. In our model, MW satellite considerations provide a practically identical bound on $\alpha$ to those of Lyman-$\alpha$ data.

Using the impact on the matter power spectrum expected in the model of nonthermal dark matter we have presented here, we have constrained the physical parameters of our model: the velocity imparted at the dark matter production, characterized by the Lorentz factor $\gamma_D$, and the temperature at reheating, $\trh$. Constraints in this parameter space also inform the allowed value of $f$, the fraction of the decaying component's energy allocated to the dark matter, that is required to obtain the correct relic abundance in the absence of dark matter annihilations.  While naturalness would suggest a value of $f\sim 0.5$, our constraints show that $f$ must be less than $\sim 10^{-4}$, implying that annihilations must be considered to avoid finely tuning $f$. Our limits within the parameter space of $\trh$ and $\gamma_D$ can also be equivalently viewed as limits on the scalar decay rate $\Gamma_\phi$ [see Eq. (\ref{eq:deftrh})] and the mass hierarchy between the scalar parent and daughter dark matter particles for a two-body decay $(m_\phi = 2\gamma_D m_\chi)$.

There are many opportunities for extensions to our model. We have assumed here that all dark matter particles are born from the decay process with the same velocity, though this need not necessarily be the case. Including a range of possible velocities could tighten or relax our bounds, depending on the exact distribution of the imparted velocity.  We have also assumed that any annihilations take place via $s$-wave processes. If annihilations occur preferentially for faster particles, this could shift the peak of our velocity distribution to lower velocities. Finally, we have only considered the cooling of dark matter due to the redshifting of its momentum. If dark matter is allowed to exchange momentum with Standard Model particles, this could provide an additional mechanism to reduce its momentum and lower the peak velocity of the velocity distribution, perhaps allowing for the formation of microhalos from perturbations that grow linearly during the EMDE. We leave these investigations for future work.

\acknowledgements
The authors would like to thank Kimberly Boddy, Julian Mu$\tilde{\mathrm{n}}$oz, Jessie Shelton, and Matteo Viel for useful discussions. This work made use of the Ulysses SISSA/ICTP supercomputer. C.M. and A.L.E. were partially supported by NSF Grant No. PHY-1752752. R.M. is partially supported by the INFN-INDARK PD51 grant.

\end{document}